\documentclass[floatfix,aps,prl,twocolumn,noeprint,superscriptaddress,notitlepage]{revtex4-1}
\usepackage{graphicx} 
\usepackage{epsfig}
\usepackage{epstopdf}
\usepackage{amsmath}
\usepackage{amsfonts}
\usepackage{amssymb} 
\usepackage{xcolor}
\usepackage{multirow}
\usepackage{lineno}
\usepackage{url}

\begin{document}
\nolinenumbers

\title{Deeply virtual Compton scattering cross section at high Bjorken $x_B$}
\author{	F.~Georges	} \affiliation{Université Paris-Saclay, CNRS/IN2P3, IJCLab, 91405 Orsay, France}
\author{	M.N.H.~Rashad	} \affiliation{	Old Dominion University, Norfolk, Virginia 23529, USA	}	
\author{	A.~Stefanko	} \affiliation{	Carnegie Mellon University, Pittsburgh, Pennsylvania 15213, USA	}
\author{	M.~Dlamini	} \affiliation{	Ohio University, Athens, Ohio 45701, USA	}
\author{	B.~Karki	} \affiliation{	Ohio University, Athens, Ohio 45701, USA	}	
\author{	S.F.~Ali	} \affiliation{	Catholic University of America, Washington, DC 20064, USA}	
\author{	P-J.~Lin	} \affiliation{Université Paris-Saclay, CNRS/IN2P3, IJCLab, 91405 Orsay, France}
\author{	H-S~Ko	} \affiliation{Université Paris-Saclay, CNRS/IN2P3, IJCLab, 91405 Orsay, France}\affiliation{Seoul National University, 1 Gwanak-ro, Gwanak-gu, 08826 Seoul, Korea}
\author{	N.~Israel	} \affiliation{	Ohio University, Athens, Ohio 45701, USA	}
\author{	D.~Adikaram} \affiliation{	Thomas Jefferson National Accelerator Facility, Newport News, Virginia 23606, USA	}			
% ...
\collaboration{The Jefferson Lab Hall A Collaboration}

\date{\today}

\begin{abstract}
We report high-precision measurements of the Deeply Virtual Compton Scattering (DVCS) cross section at high values of the Bjorken variable $x_B$. DVCS is sensitive to the Generalized Parton Distributions of the nucleon, which provide a three-dimensional description of its internal constituents.

Using the exact analytic expression of the DVCS cross section for all possible polarization states of the initial and final electron and nucleon, and final state photon, we present the first experimental extraction of all four helicity-conserving Compton Form Factors (CFFs) of the nucleon as a function of $x_B$, while systematically including helicity flip amplitudes. In particular, the high accuracy of the present data demonstrates sensitivity to some very poorly known CFFs.  

\end{abstract}


\maketitle


  {    
In this letter, we present a comprehensive experimental determination of the 12 complex helicity amplitudes of the $\gamma^\ast p \to \gamma p$ amplitude, measured in the deeply virtual Compton scattering (DVCS) reaction $ep\to ep\gamma$.  This amplitude is illustrated in Fig.~\ref{figure1}, which also defines our kinematic nomenclature.
The Bjorken limit of DVCS, first described in \cite{Ji:1996ek}, is defined by large virtuality $Q^2$ and large invariant `energy' $\nu = q\cdot P/M$ of the virtual photon   at fixed $x_B = Q^2/(2q\cdot P)$
and small net momentum transfer to the proton. QCD theorems~\cite{Collins:1996fb,Ji:1998xh}  prove the DVCS amplitude  factorizes into a hard perturbative kernel and a soft part described by light cone matrix elements \cite{Mueller:1998fv} of quark and gluon operators.  In this scaling limit, the $\gamma^\ast p\to\gamma p$ amplitude reduces to just 4 complex amplitudes, whose $Q^2$-dependence
is determined by QCD evolution equations~\cite{Balitsky:1997mj}.
The light cone matrix elements, also called Generalized Parton
Distributions (GPDs), encode tomographic images correlating the transverse spatial and longitudinal momentum distributions of quarks and gluons inside the proton, leading to a sum rule for the separate contributions of quarks and gluons to the spin of the proton~\cite{Ji:1996ek}.

The $ep$ scattering kinematics in the Bjorken limit define a preferred longitudinal axis (up to ambiguities of order $t/Q^2$). Light cone momenta $P^\pm = (P^0\pm P^z)/\sqrt{2}$ and light cone helicities of the external particles are defined with respect to this axis.  The variables $x\pm \xi$ are the light cone momentum fractions of the initial and final active quark.  The variable $\xi$ is kinematic: $\xi \approx x_B/(2-x_B)$, whereas $x$ is integrated
from $-1$ to $1$ as a consequence of the implied quark loop. The experimental 
$e p\to e p\gamma$ scattering amplitude is the coherent sum of the Compton
amplitude and the  Bethe-Heitler (BH) amplitude, wherein the real photon is emitted by the incoming or the scattered electron, as illustrated in Fig.~\ref{figure1}

\begin{figure}[b]
  \centering\includegraphics[width=\linewidth]{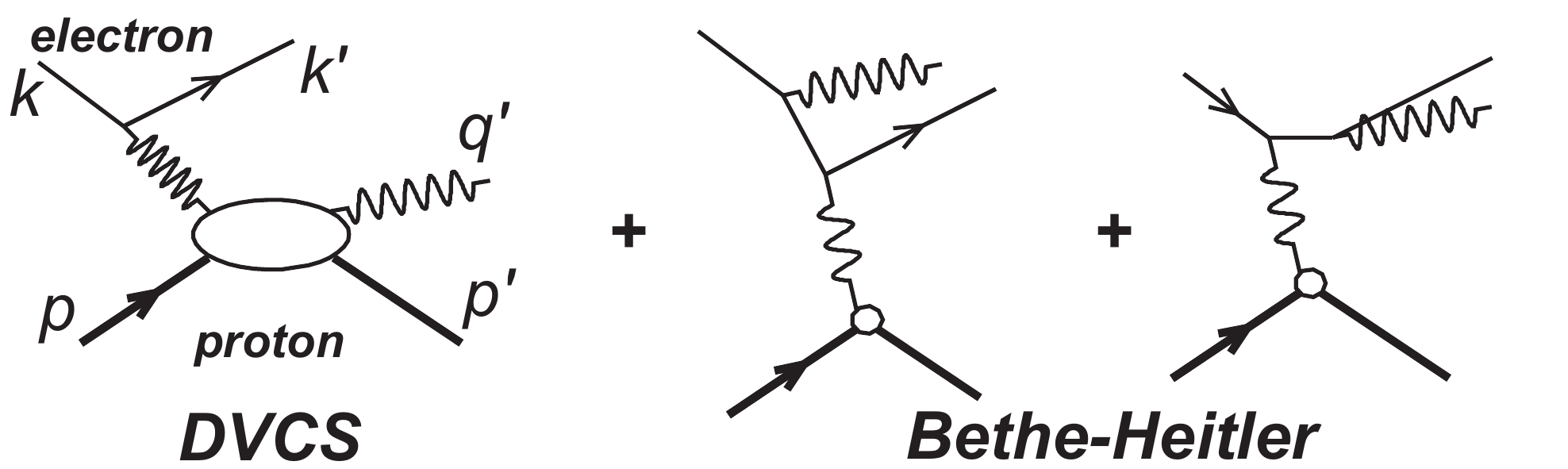}
\caption{ 
Lowest-order QED diagrams for the process $ep\rightarrow ep\gamma$,
including the DVCS and Bethe-Heitler (BH) amplitudes. The external
momentum four-vectors are defined on the diagram. The virtual photon
momenta are $q=k-k'$ in the DVCS- and $\Delta=q-q'$ in the BH-amplitudes.
The invariants are: $W^2=(q+p)^2$, $Q^2=-q^2>0$, $t=\Delta^2$, $x_{B}=Q^2/(2 p\cdot q)$, and the DVCS scaling variable
$\xi=-\overline{q}^2/(\overline{q}\cdot P)\approx x_{B}/(2-x_{B})$, with $\overline{q}= (q+q')/2$ and $P=p+p'$.
}
  \label{figure1}
\end{figure}

In this analysis of the Jefferson Lab Hall A experiment E12-06-114, we follow the BMMP formalism  \cite{Braun:2014}, wherein the longitudinal axis  is defined in
an event-by-event frame in which the three-vectors $\mathbf q$ and $\mathbf q'$ are colinear.  More generally, the light cone is defined by null vectors $q'$ and
$q-q'/(1-t/Q^2)$. In this reference frame, the leading  four Compton amplitudes
conserve  the light cone helicity of the photons.
The proton helicity dependence of the Compton amplitude is expressed through the definition of
four chiral-even  Compton form factors (CFFs) ($\mathcal H_{++}$, $\widetilde{\mathcal H}_{++}$, $\mathcal E_{++}$, $\widetilde{\mathcal E}_{++}$), which are convolution integrals of the four corresponding GPDs.  Each CFF is associated with
a unique nucleon-spinor matrix element of \textit{e.g.}
$\gamma^+,\ \gamma^+\gamma_5,\ldots$

The reduction of the twelve Compton amplitudes to
just four amplitudes, as first described in \cite{Ji:1998xh}
is a profound simplification.  Nonetheless in the range
of $Q^2$ and $x_B$ currently accessible, the remaining eight chiral-odd photon helicity-flip Compton amplitudes, while small, cannot be completely neglected.

The HERMES collaboration performed extensive measurements of single- and double-spin DVCS asymmetries \cite{Airapetian:2012pg,Airapetian:2010ab,Airapetian:2011uq}.  The H1
\cite{Aaron:2007ab} and ZEUS \cite{Chekanov:2008vy} collaborations measured the
DVCS cross section over a broad range of $Q^2$ and $W^2$ at low $x_B$.
The Jefferson Lab CLAS collaboration has measured
DVCS beam spin asymmetries and cross sections \cite{Girod:2007aa,Jo:2015ema,HirlingerSaylor:2018bnu} and longitudinally-polarized target asymmetries
\cite{Chen:2006na,CLAS:2014qtk,CLAS:2015bqi}. Recent experimental studies on DVCS show that the contributions of the chiral-even GPDs dominate the DVCS amplitude  
already  at $Q^2$ values as low as 1.5 GeV$^2$~\cite{MunozCamacho:2006hx,Defurne:2015kxq,Jo:2015ema}.
However, dynamic terms involving a photon helicity flip are not negligible,
even though they are nominally suppressed by powers of $(t, M^2)/Q^2$ \cite{Defurne:2017paw}.

\begin{table*}[!t]
  \begin{tabular}{|c||c|c|c||c|c|c|c||c|c|}
    \hline
    Setting & Kin-36-1 & Kin-36-2 & Kin-36-3 & Kin-48-1 & Kin-48-2 & Kin-48-3 & Kin-48-4 & Kin-60-1 & Kin-60-3\\
    \hline
    $x_B$ & \multicolumn{3}{|c||}{0.36} & \multicolumn{4}{|c||}{0.48} & \multicolumn{2}{|c|}{0.60} \\
    \cline{2-10}
   $E_b$ (GeV) & 7.38 & 8.52 & 10.59 & 4.49 & 8.85 & 8.85 & 10.99 & 8.52 & 10.59 \\
    $Q^2$ (GeV$^2$) & 3.20 & 3.60 & 4.47 & 2.70 & 4.37 & 5.33 & 6.90 & 5.54 & 8.40 \\
    $E_\gamma$ (GeV) & 4.7 & 5.2 & 6.5 & 2.8 & 4.7 & 5.7 & 7.5 & 4.6 & 7.1 \\
    $-t_{min}$ (GeV$^2$) & 0.16 & 0.17 & 0.17 & 0.32 & 0.34 & 0.35 & 0.36 & 0.66 & 0.70 \\
    $\int Q\,dt$ (C) & 1.2 & 1.7 & 1.3 & 2.2 & 2.2 & 3.7 & 5.7 & 6.4 & 18.5 \\
    \hline
    $\#$ data bins & \multicolumn{3}{|c||}{672} & \multicolumn{4}{|c||}{912} & \multicolumn{2}{|c|}{480} \\
    \hline
  \end{tabular}
  \caption{Main kinematic variables for each of the nine ($Q^2, x_B$) settings where the DVCS cross section is reported. $E_b$ is the incident electron energy, $E_\gamma$ and $-t_\text{min}$ correspond to a final state photon emitted parallel to $\mathbf q = \mathbf k-\mathbf k'$ at the nominal
  $Q^2$, $x_B$ values listed.
  For each setting, the cross section is measured as a function of $t$ (3 to 5 bins depending on the setting) and in 24 bins in $\phi$. The accumulated charge, corrected by the acquisition dead-time, is listed in the row labeled $\int Q dt$. The last row of the table indicates the number of statistically independent measurements (bins) for each $x_B$ setting, including helicity-dependence.}
  \label{tab:kin}
\end{table*}

This letter reports the results of experiment E12-06-114, which ran in Hall A at Jefferson Lab in the fall of 2014 and in 2016.
Concurrent data on $ep\to ep\pi^0$ were published in \cite{JeffersonLabHallA:2020dhq}, which also includes additional experimental and analysis details.  Table~\ref{tab:kin} shows the nine kinematic settings in $Q^2$ and $x_B$ at which the  DVCS cross sections were measured. For each setting, the data are binned in $t$ and the azimuth $\phi$ of the detected photon $q'$ around the direction of $\mathbf q$, as defined by the ``Trento convention'' \cite{Bacchetta:2004jz}.

The longitudinally polarized electron beam impinged on a 15-cm liquid hydrogen target. The beam current was adjusted between 5 and 15 $\mu$A, depending on the kinematic setting, in order to maintain dead-time below 5\%.   The  Hall A M\o ller polarimeter measured an averaged beam polarization of  86$\pm 1\%$.  The $H(\vec{e},e' \gamma)X$ reaction was the main trigger of the data acquisition system.  The scattered electron was detected by a coincidence signal between the scintillators and the Cerenkov detector of the Left High-Resolution Spectrometer (HRS)~\cite{Alcorn:2004sb}. The electron identification was further refined offline by the use of a Pb-Glass calorimeter in the HRS.  The outgoing photon was detected by a dedicated highly-segmented PbF$_2$ electromagnetic calorimeter. The analog signal
from each of the 208 calorimeter channels was recorded over 128~ns by 1~GHz digitizing electronics based on the Analog Ring Sampler (ARS) chip~\cite{Feinstein:2003vi,Druillole:2001dm}.  Following an HRS electron trigger (level-1), calorimeter signal sampling was stopped. Waveform digitization was validated by a level-2 trigger which computed the sum of the signal from all channels in a 80~ns window. If a signal above a programmable threshold was found in the calorimeter, the digitization process took 128 $\mu$s; otherwise the ARS system resumed sampling after 500~ns. The level-2 trigger was based on a field-programmable gate array (FPGA) module, and was used only during high counting rate settings ($>1$~kHz). For settings with low rates, all level-1 triggers were validated and waveforms digitized~\cite{JeffersonLabHallA:2020dhq}. Offline analysis of the calorimeter signals and regular energy calibrations resulted in an energy resolution of 3\%   at 7 GeV.   Missing-mass reconstruction identified the non-detected proton  (see Fig.~\ref{fig:mm}).  The time resolution between the electron and photon detections was better than 1~ns. The number of random coincidences was evaluated by analyzing events in a time window shifted with respect to the coincidence time of the HRS and calorimeter signals.
 
\begin{figure}[tb]
  \includegraphics[width=\linewidth]{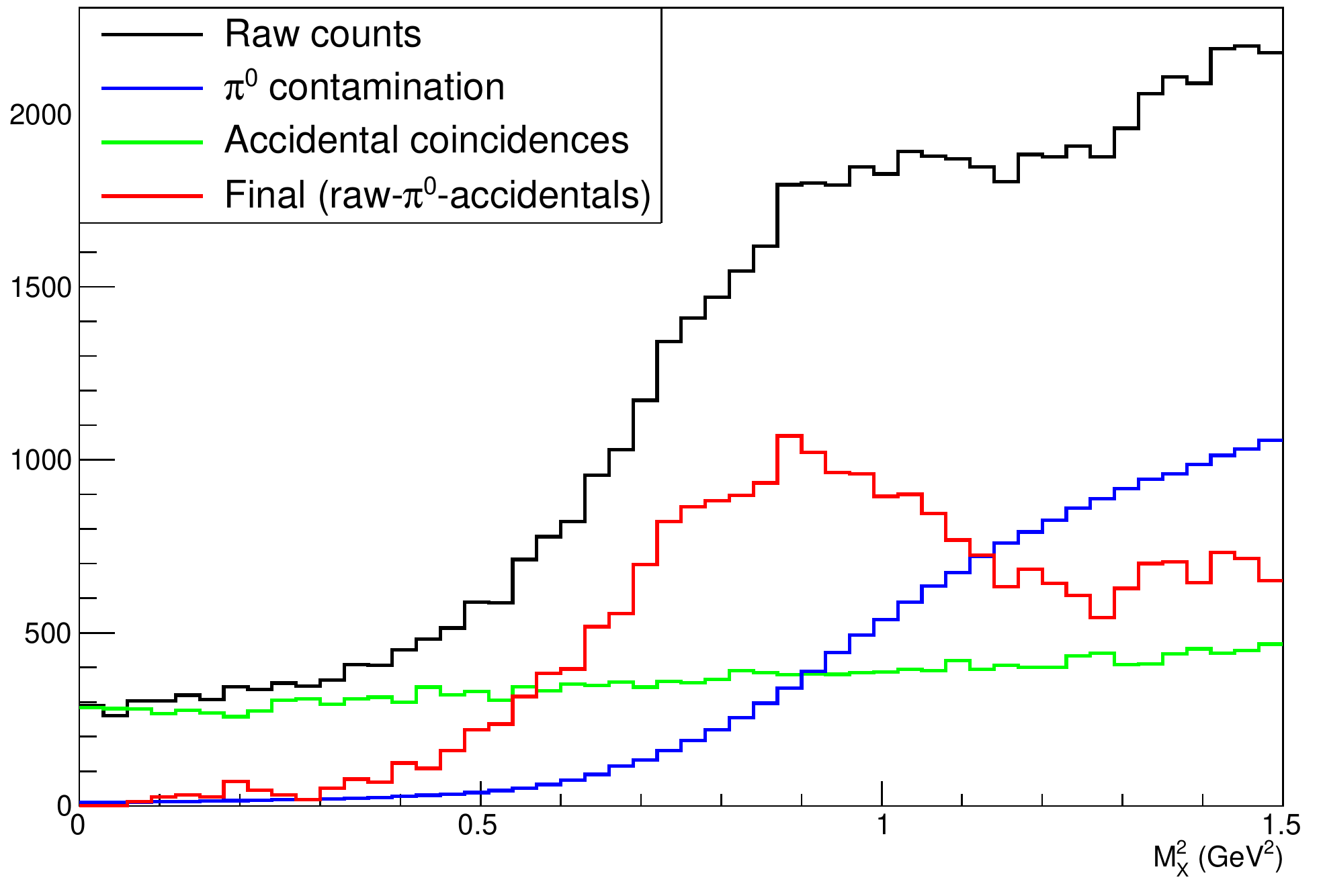}
  \caption{Missing mass squared of the $ep\to e\gamma X$ reaction for kinematic setting Kin-48-1, integrated over $t$ and $\phi$. Experimental  data are shown in black. The subtraction of the accidental contribution (green) and photons from $\pi^0$ decays (blue) yields the red histogram. }
  \label{fig:mm}
\end{figure}

An important source of background was neutral-pion electroproduction events for which only one of the decay photons was detected. The number of one-photon events from $\pi^0$ decays was estimated by a Monte Carlo simulation normalized bin by bin to the number of detected $\pi^0 \to \gamma\gamma$ events. The acceptance and resolution of the experiment were modeled by a {\sc Geant4} simulation.  The simulation included bin migration effects due to real and virtual radiation and the PbF$_2$ calorimeter energy resolution, as described in \cite{Defurne:2015kxq}.  During the data taking, the first quadrupole of the HRS experienced the gradual failure of its cryogenic current lead. For the first part of the experiment, the faulty quadrupole could only be used at a reduced current supply. Before the Fall 2016 data taking, that quadrupole was replaced by a room-temperature quadrupole providing a similar magnetic field. Optics calibrations data were taken to maintain the excellent resolution of the HRS. Effects on the spectrometer acceptance were taken into account for each kinematic setting and run-period by applying similar multidimensional cuts (R-cuts, \cite{Rvachev:2001}) on both the experimental and simulated data.  

DIS data were taken simultaneously to the main DVCS data using an ancillary trigger for all kinematic settings, which allowed a monitor of the scattered electron detection efficiency and acceptance~\cite{JeffersonLabHallA:2020dhq}. 
The total systematic uncertainty of the DVCS cross-section measurements includes the uncertainty on the electron detection and acceptance, the luminosity evaluation, the uncertainty on the
photon detection, and the exclusivity. Radiative corrections are included in the analysis based on calculations of~\cite{Vanderhaeghen:2000ws} and using the procedure described in detail in~\cite{JeffersonLabHallA:2020dhq}.

\begin{figure*}
\includegraphics[width=\linewidth]{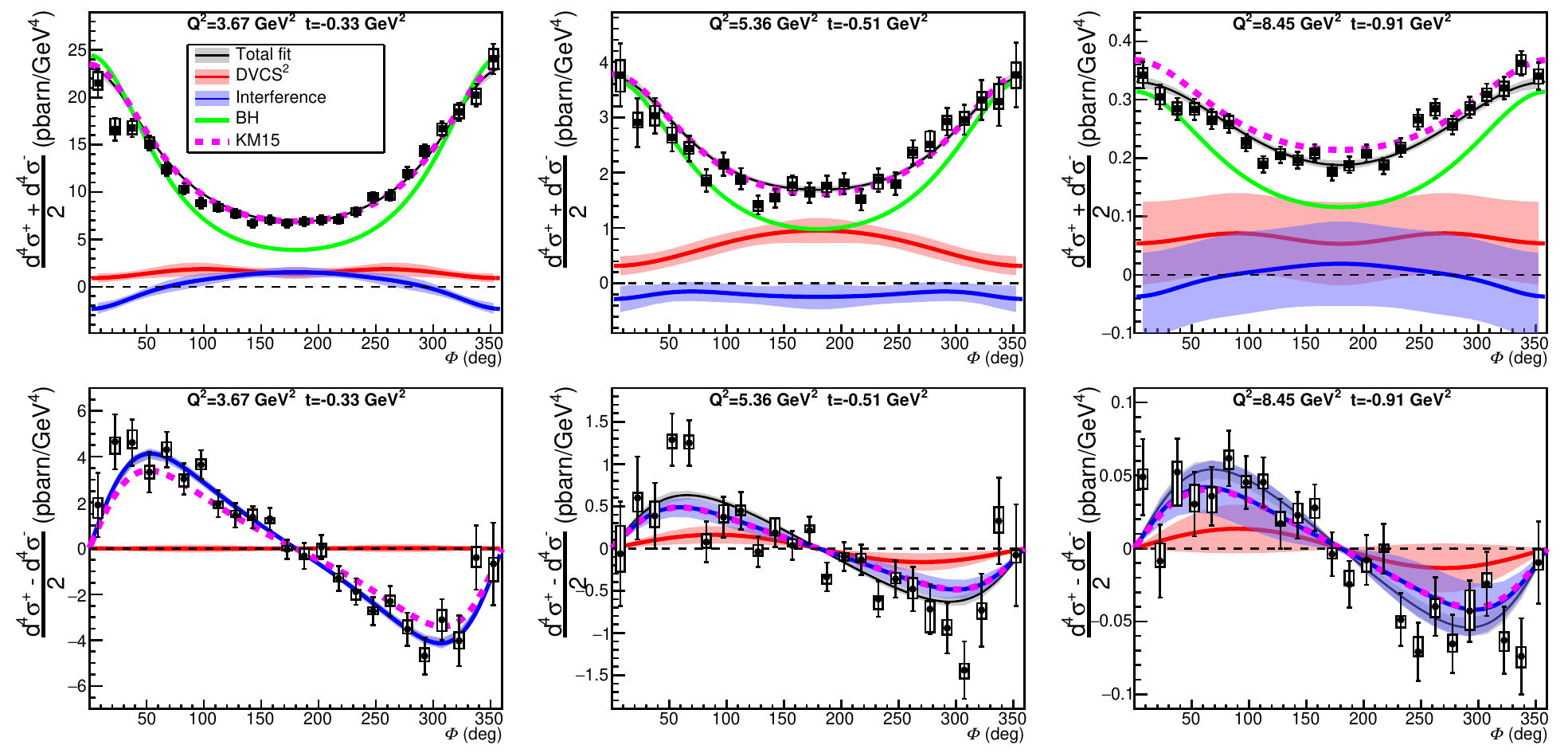}
\caption{Helicity-independent (top) and helicity-dependent (bottom) DVCS cross
cross-section at $x_B=0.36$ (left), $x_B=0.48$ (center) and $x_B=0.60$ (right)
for the values of $Q^2$ and $t$ indicated on the top of each figure. Bars around the points indicate statistical uncertainty and boxes show the total systematic uncertainty, computed as the quadratic sum of the point-to-point and correlated systematic uncertainties. Black
curves display the total fit to the cross sections, at constant $x_B$ and $t$, in
the BMMP formalism. The BH cross section is shown in green. The contribution
from the BH-DVCS interference is shown by the blue bands, whereas the
contribution from the DVCS$^2$ term is indicated by the red bands. All band
widths correspond to one standard deviation. The KM15 model is shown in magenta.
}
\label{fig:xsec}
\end{figure*}

Figure~\ref{fig:xsec} shows a sample of the cross section measured at each of the $x_B$ settings. The azimuthal dependence of the cross section is fit using the BMMP formalism~\cite{Braun:2014}, and the contribution from the BH-DVCS interference and DVCS$^2$ contributions are shown along with the BH cross section. The BMMP calculation includes kinematic power corrections $\sim t/Q^2$ and $\sim M^2/Q^2$ that were proven to be important at these kinematics~\cite{Defurne:2017paw}. The cross section is expressed as a function of helicity-conserving CFFs ($\mathcal H_{++}$, $\widetilde{\mathcal H}_{++}$, $\mathcal E_{++}$, $\widetilde{\mathcal E}_{++}$), longitudinal-to-transverse helicity-flip CFFs ($\mathcal H_{0+}$, $\widetilde{\mathcal H}_{0+}$, $\mathcal E_{0+}$, $\widetilde{\mathcal E}_{0+}$) and transverse helicity-flip CFFs ($\mathcal H_{-+}$, $\widetilde{\mathcal H}_{-+}$, $\mathcal E_{-+}$, $\widetilde{\mathcal E}_{-+}$). For each GPD label, the subscripts $\lambda,\lambda'$ refer to the light cone helicity of the initial (virtual) and final (real) photon, respectively. In this formalism, the light cone is defined by linear combinations of $q^\mu$ and $q^{\prime\mu}$. Our whole dataset has been fitted using this complete and consistent scheme, with the real and imaginary part of all these CFFs being the free parameters (a total of 24) of the fit. All kinematics bins  in $Q^2$ and $\phi$ at constant $(x_B,t)$ are fitted simultaneously, however possible QCD evolution
of the CFFs as functions of $Q^2$ is not considered.

While the number of fit parameters is large, the high accuracy of the data allows to simultaneously extract all the helicity-conserving CFFs with good statistical uncertainties. Figure~\ref{fig:cff} shows the real and imaginary part of all 4 helicity-conserving CFFs as a function of $x_B$ averaged over $t$. These results represent the first complete extraction of all CFFs appearing in the DVCS cross section, including the poorly known $\mathcal E_{++}$ and $\widetilde{\mathcal E}_{++}$. The state-of-the-art GPD parametrization KM15~\cite{Kumericki:2015lhb} that reproduces worldwide DVCS data show a reasonable agreement but fail to describe $\mathcal E_{++}$ and $\widetilde{\mathcal E}_{++}$ accurately.

As first demonstrated in~\cite{Defurne:2017paw} and described theoretically in~\cite{Kriesten:2019jep},
the measurement of the DVCS cross section
at two or more values of the $ep$ center-of-mass energy $\sqrt{s}$ provides statistically significant
separation of the 
real and imaginary parts of the BH-DVCS interference term as well as the DVCS$^2$ contribution 
in the cross sections for polarized electrons.
A new analysis~\cite{Cuic:2020iwt} of all previous JLab DVCS data followed a similar
procedure, and obtained flavor-separated Compton Form Factors,
after inclusion of  our
recent neutron DVCS data~
\cite{Benali:2020vma}.
In the present analysis,
realistic error bands on the chiral-even CFFs are obtained by explicit inclusion of higher-order terms (\textit{e.g.} $\mathcal H_{0+},\, \mathcal H_{-+}$,\textit{etc.}) in the cross section fit, with these
terms primarily constrained by inclusion
of higher Fourier terms in the azimuthal variable $\phi$.  
Although the extracted values of the helicity-flip CFFs are largely statistically consistent with zero, the statistical correlations between all of the CFF
values at fixed $x_B$ are essential to obtaining
realistic experimental uncertainties.  

The sensitivity to
the CFFs $\mathcal E$ and $\widetilde {\mathcal E}$
illustrated in Fig.~\ref{fig:cff} arises from the $Q^2$-dependent kinematic
factors weighting these terms relative to the
contributions of $\mathcal H$ and $\widetilde{\mathcal H}$.
The KM15 model~\cite{Kumericki:2015lhb} includes
only the $D$-term (support limited to $|x|<\xi$)
in the $E$ GPD, and therefore vanishes at $x=\xi$,
resulting in $\mathcal I\text{m}[\mathcal E]=0$.  For $\widetilde E$,
this model includes only the pion pole, via the
$\gamma*\gamma\to\pi^0$ amplitude, and thus the
amplitude in this channel is also purely real.
In contrast, the model of \cite{Goeke:2001} for $E$ includes 
a valence quark contribution with support outside
the $|x|<\xi$ bound and therefore produces a non-zero imaginary part of the $\mathcal E$ CFF. 
Similarly, the chiral quark soliton model \cite{Penttinen:1999th,Goeke:2001}
produces a contribution to $\widetilde E$ that while smaller in magnitude to the pion-pole, is  additive with opposite sign.  This may explain the significant
difference between our values of 
$\mathcal R\text{e}[\widetilde{\mathcal E}]$
and the KM15 model.
GPDs can be described as momentum decompositions of the corresponding form factors.  This is explicit
in the  first moment sum rules, which relate \textit{e.g.} GPDs $E$ and $\tilde E$ (summed over  quark flavor $f$) to
the axial and pseudo-scalar form factors $G_A$ and
$G_P$ of the proton:
\begin{align}
\sum_f \int_{-1}^1 
\left\{\begin{array}{c}E_f(x,\xi,t) \\ \widetilde E_f(x,\xi,t) \end{array}
\right\}
dx &= \left\{\begin{array}{c} G_{A}(-t) \\ G_P(-t)
\end{array}\right\}
\end{align}
These form factors, particularly $G_P$ are much less well known experimentally than the usual electromagnetic form factors $G_{E,M}$.
The present measurements of the CFFs $\mathcal E$
and $\widetilde{\mathcal E}$ therefore provide constraints on the quark momentum distribution support
of the corresponding form factors within this
$x_B$ range.

The present measurements will be complemented
in this same general kinematic range in the near future by measurements in JLab Halls B and C,
and longitudinally polarized proton measurements
 and neutron DVCS measurements
 in JLab Hall B.
 These measurements therefore demonstrate
that the full extraction of experimental
Compton form factors is within reach.

\begin{figure}[hbt]
\includegraphics[width=\linewidth]{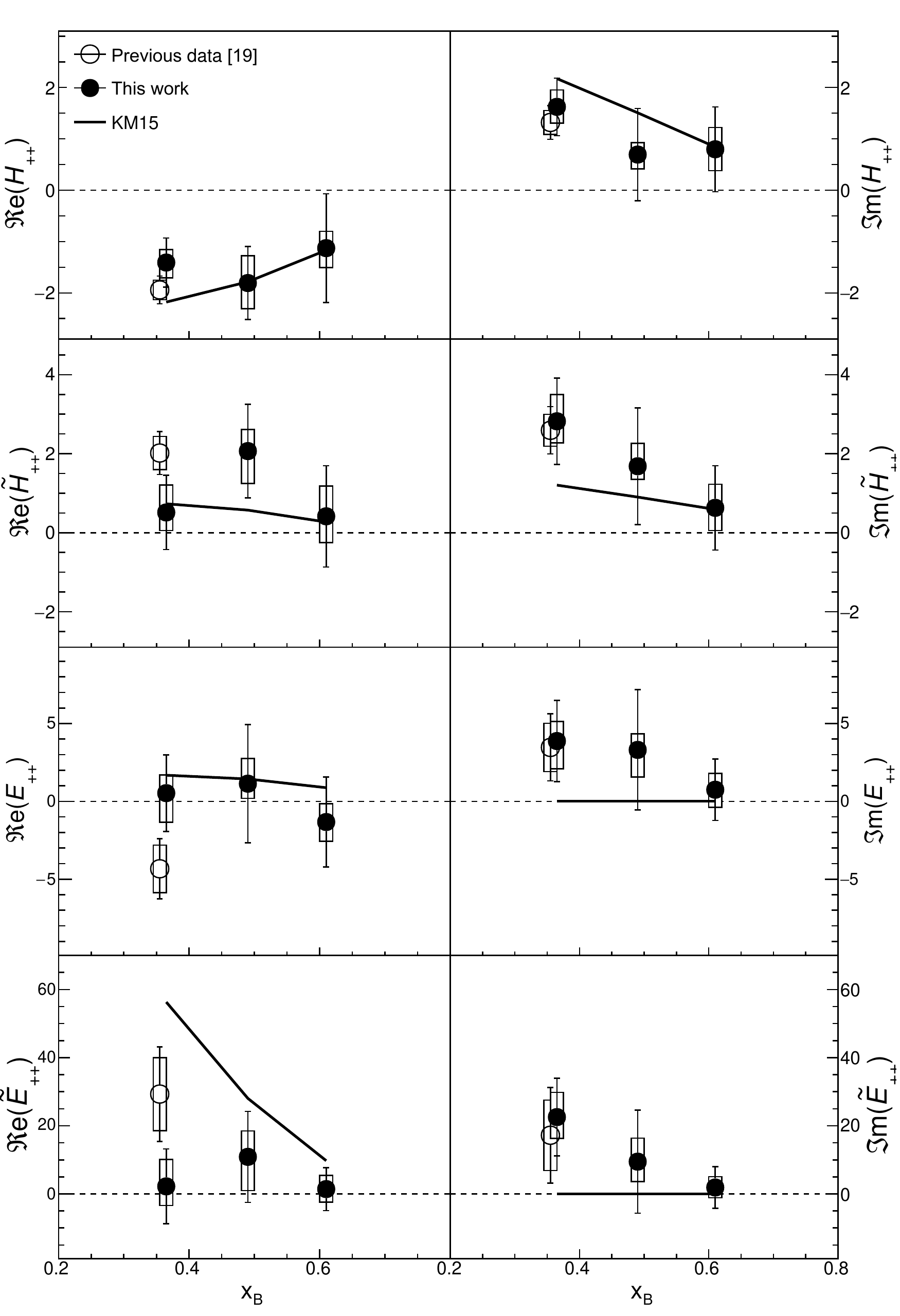}
\caption{Values of the helicity-conserving CFFs, averaged over $t$, as a function of $x_B$. Bars around the points indicate statistical uncertainty and boxes show the total systematic uncertainty.
The fit results of previous data~\cite{Defurne:2015kxq} at $x_B=0.36$ are displayed with the open
markers. The average $t$ values are $-0.281$ GeV$^2$ ~\cite{Defurne:2015kxq} and $-0.345$, $-0.702$,  $-1.050$ GeV$^2$ at $x_B=0.36, 0.48,\, 0.60$, respectively.  The solid lines show the KM15 model~\cite{Kumericki:2015lhb}.}
\label{fig:cff}
\end{figure}

\section*{Acknowledgements}

We acknowledge essential contributions by the Hall A collaboration and Accelerator and Physics Division staff at Jefferson Lab. This material is based upon work supported by the U.S. Department of Energy, Office of Science, Office of Nuclear Physics under contract DE-AC05-06OR23177. This work was also supported by a DOE
Early Career Award to S. Covrig Dusa for  the development of the
high power hydrogen target cells,
the National Science Foundation (NSF), the French CNRS/IN2P3, ANR,  and P2IO Laboratory of Excellence, and the Natural Sciences and Engineering Research  Council of Canada (NSERC).  We also wish to acknowledge the many contributions to this subject,
and the personal encouragement of our friend,
Maxim Polyakov, recently deceased.
}
\bibliography{Compton2018}

\end{document}